\documentclass[10pt,conference,letterpaper]{RWWTemplate}

\usepackage{times,amsmath,epsfig}
\usepackage{cite}
\usepackage{amsmath,amssymb,amsfonts}
\usepackage{algorithmic}
\usepackage{gensymb}
\usepackage{subfigure}
\usepackage{graphicx}
\usepackage{textcomp}

\usepackage{multirow}
\usepackage{multicol}

\usepackage{tablefootnote}
\usepackage{booktabs}
\usepackage{textgreek}
\usepackage{threeparttable}
\usepackage[utf8]{inputenc}

\usepackage{xcolor}

\renewcommand{\baselinestretch}{0.87}

\begin{document}

\title{Compact-Range RCS Measurements and Modeling of Small Drones at 15 GHz and 25 GHz\\
}

\author{%
Martins Ezuma, Mark Funderburk, and Ismail Guvenc\\
Dept. Electrical and Computer Engineering, North Carolina State University, Raleigh, NC 27606\\
\{mcezuma, mtfunder, iguvenc\}@ncsu.edu)}%


\maketitle

\begin{abstract}
The knowledge of the radar signature of aerial targets, such as drones, is critical in designing an effective radar detection system. It is a challenging task to measure the radar cross-section (RCS) of small drones. 
This paper describes a compact-range approach for measuring the RCS of small drones at 15 GHz and 25 GHz. The measurement results show that the average RCS of the three small drones varies with the radar frequency with higher reflections observed around certain directions. Moreover, the results show that for each drone, the RCS at 25 GHz is higher than the RCS at 15 GHz. Besides, information-theoretical based model selection for the RCS data is carried using the Akaike information criterion (AIC). We find that the generalized extreme value distribution is a good fit for modeling the RCS of small drones. 
\end{abstract}


\section{Introduction}

In recent times, small drones are finding applications in precision agriculture, remote sensing, delivery of goods, search and rescue missions, law enforcement, and infrastructure inspections, among others~\cite{shakhatreh2018unmanned}. Low altitude drones equipped with high-resolution thermal cameras and LiDAR/magnetometer sensors are used to generate 3D geophysical maps which are quickly replacing ground survey maps. Judging by the current trend in drone applications, it is expected that small drones will become increasingly popular in the future. This is not all good news, because, in recent times, drones have been used to carry out crimes and terror attacks, and may threaten public safety if used by malicious entities.



Existing techniques for the detection of unauthorized drones include radar-based, radio frequency-based, acoustic-based, vision-based, and sensor fusion based techniques~\cite{ezuma2019micro,guvenc2018detection}. Each of these techniques has its advantages and disadvantages. Radar-based techniques are appealing since radars operate in different weather conditions. Moreover, radars can be easily deployed on different platforms: land, sea, and air.
 
  Active radars can detect objects by transmitting electromagnetic waves and listening for the echo (back-scattered signals) from the targets. Therefore, the ability of a radar to detect a particular target depends on the radar reflective/scattering properties of the targets. This scattering property of a specific target is termed the radar cross-section (RCS) of the target. Targets with higher RCS are more visible to a given radar than targets with lower RCS. The RCS of a target depends on the frequency of the radar, target shape, aspect angle and material design of the target. Studies have suggested that small drones, by design, have very low RCS~\cite{guvenc2018detection,MartinsDrone}. This explains why it is difficult for many traditional radars to detect a small drone. In~\cite{White_House}, a White House surveillance radar system, designed to detect missiles and airplanes, failed to detect a small drone that flew across the fence and crashed into the South Lawn. Moreover, studies have shown that the RCS of small drones and birds are similar in certain frequencies\cite{rahman2018flight}. This RCS similarity between small drones and birds could increase the false alarm rate of a radar designed to detect small drones~\cite{gong2019interference}. Therefore, a proper understanding of the RCS properties of small drones is vital to the design of a successful radar detection system, which is also the primary motivation for this study.
 
  There are very limited indoor experimental studies of the RCS of small drones in the literature. Most of the existing studies do not provide sufficient analytical framework for the detection of small drones using information from the RCS measurement. In~\cite{nakamura2017characteristics}, the RCS of a small drone is measured in the H-H plane in an anechoic chamber at a frequency of 2.4 GHz and 24 GHz. In this measurement, the drone is placed on a turntable which rotates in the anechoic chamber and a vector network analyzer (VNA) is used to capture the scattered power from the drone. The study shows that the average RCS of the DJI Phantom 3 drone is about -13~dBsm at 24~GHz. In~\cite{roding2017fully}, the RCS of several small drones is measured indoors in the bistatic configuration in the frequency range 30.4-37.1~GHz. The study shows that the shape, design material, electric field polarization, and bistatic angle of the radar system influence the measured RCS values of small drones.  
In these indoor RCS measurement scenarios reported in the literature, the transmit and receive antennas may not be far enough from the target drone to generate plane wave illuminations of the drones. Plane waves scattering is required for accurate RCS measurement. To guarantee the generation of plane waves during indoor RCS measurement, a large parabolic reflector is required in the anechoic chamber. This chamber configuration is called a compact range. Therefore, there is a need to carry out RCS measurement of small drones in a compact range anechoic chamber arrangement.

In this study, we carry out a compact-range RCS measurement of three small drones in an anechoic chamber at 15 GHz and 25 GHz. The choice of these frequencies is motivated by the fact that several commercial radar systems designed for drone detection operate in these frequency bands~\cite{ancortek_radar,fortem_radar}. The RCS data from the compact range measurement is modeled using three different probability distributions. A distribution that can accurately model the RCS data can be used to develop an effective radar-based statistical detection system for small drones. The remainder of this paper is organized as follows. Section~\ref{rcs_measurement} describes the compact-range RCS measurement system and technique, Section~\ref{result_DISCUSSION} presents the results of the measurement and statistical modeling, and Section~\ref{conclusion_sec} presents the conclusion of the work.

\begin{figure*}{}
\center{
 \begin{subfigure}[]{\includegraphics[scale=0.18]{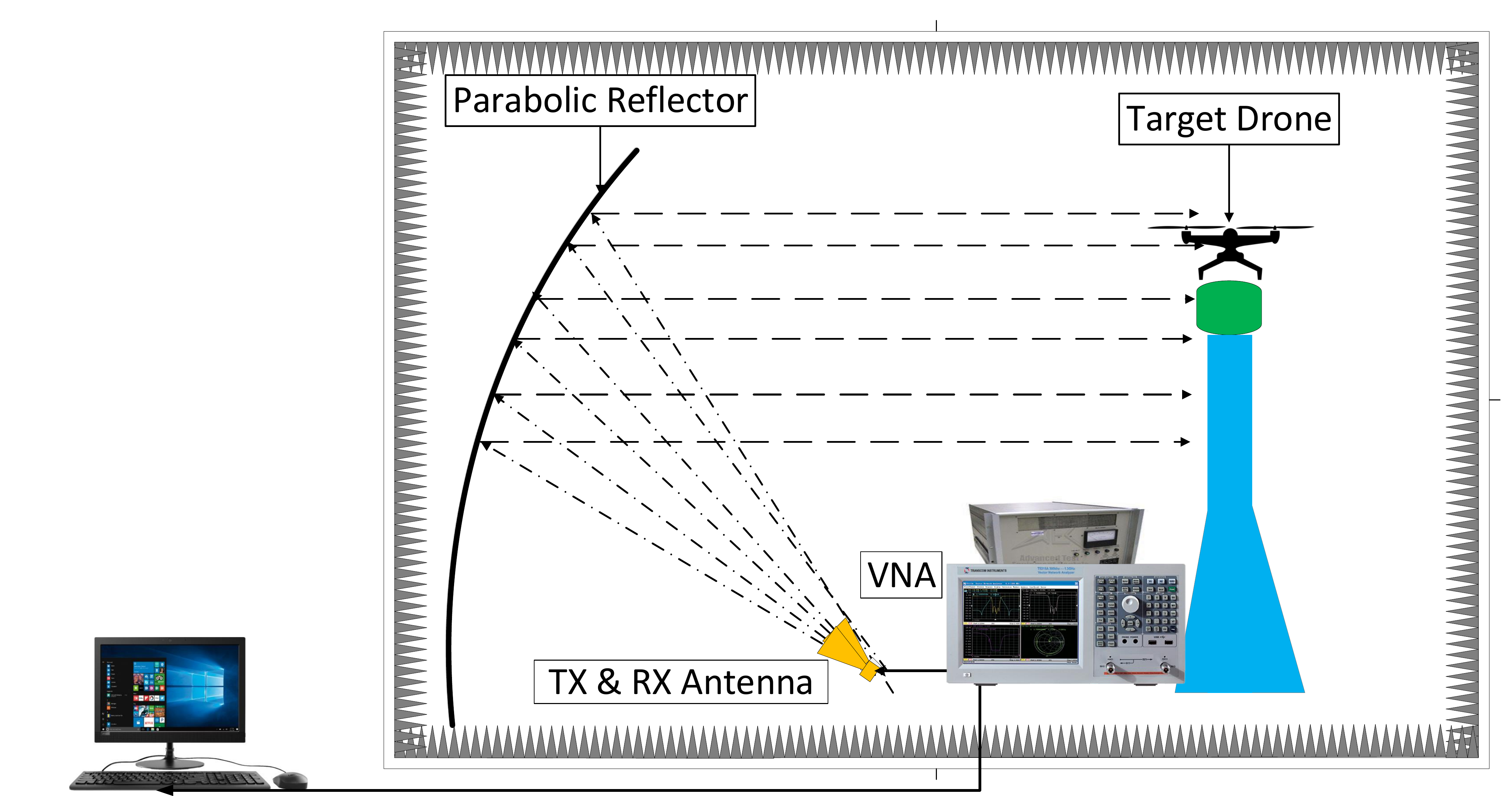}\label{EXPERIMENTAL_setup}}
\end{subfigure}
\begin{subfigure}[]{\includegraphics[width=0.3\linewidth]{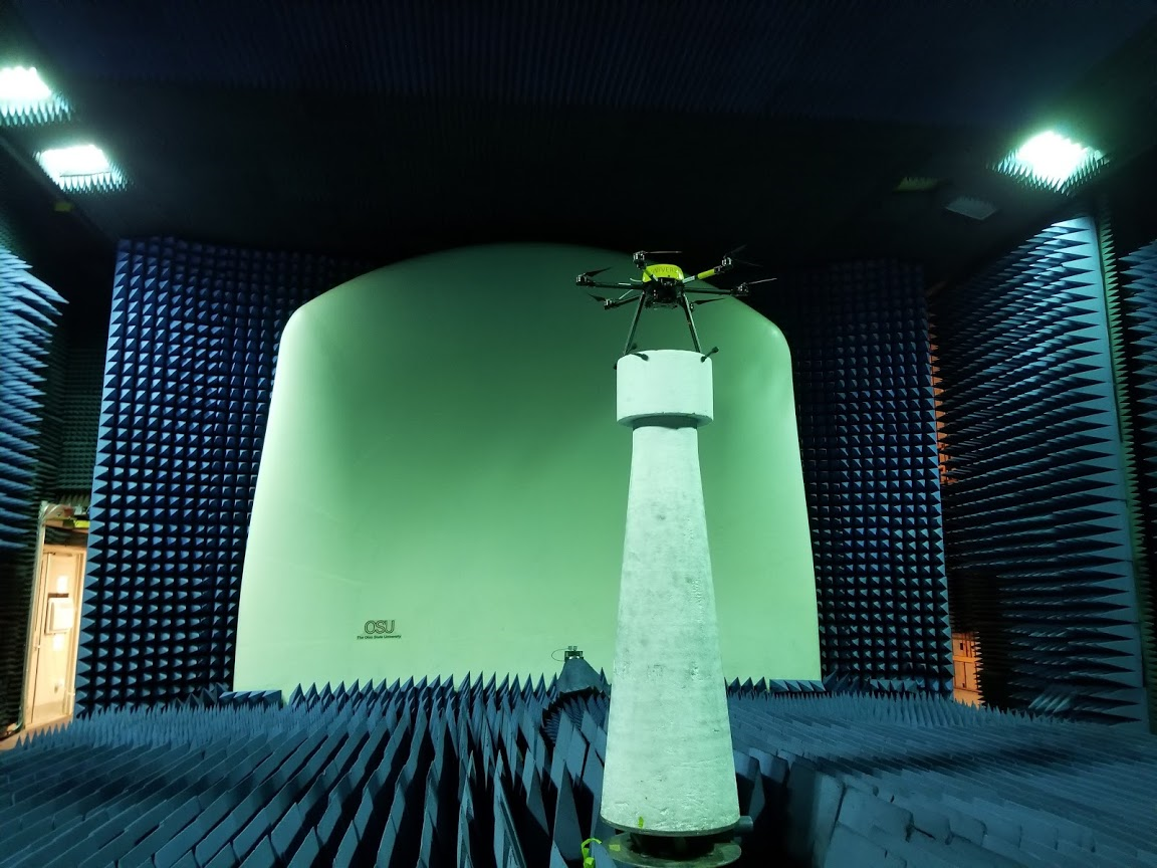}\label{ANECHOIC_CHAMBER}}
\end{subfigure}
\begin{subfigure}[]{\includegraphics[width=0.3\linewidth,height=0.23\linewidth,keepaspectratio]{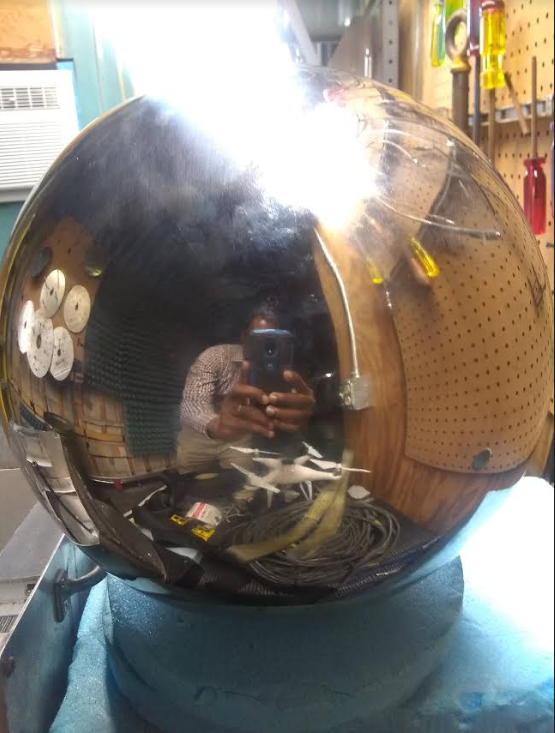}\label{caliberation_sphere}

}
\end{subfigure}
 \caption{{ (a) Schematic of the Compact-range anechoic chamber, (b) measuring the RCS of the Trimble zx5 drone, (c) the 12-inch calibration sphere used for the measurement.}}
 \label{fig:1}}

 \end{figure*}

\section{RCS Measurement}\label{rcs_measurement}
\subsection{Theoretical Background }

When a target is illuminated by radar, it scatters and diffracts the electromagnetic waves in all directions. The RCS is used to describe the amount of scattered power in the direction of the radar. If the target is in the near field of the radar, the intercepted and scattered waves have spherical wavefronts. However, if the target is in the far-field, which is a typical radar detection scenario, the wavefronts consist of a combination of plane waves. In such a case, the far-field RCS of the target is given by~\cite{knott2004radar}:
\begin{equation}
   \sigma=\lim_{R\to\infty} 4\pi R^2\frac{|E_s|^2}{|E_i|^2},
   \label{RCS_eq1}
\end{equation}
where $\sigma$ is the RCS of the target while $E_s$ and $E_i$ are the far-field scattered and incident electric field intensities respectively as seen at a distance $R$. If we model a complex target as consisting of $N$-scattering centers, then the total RCS of the target is the complex addition of the RCS of each scattering point and given by~\cite{knott2004radar}:
\begin{equation}
   \sigma=\Bigg|\sum_{i=1}^{N} \sqrt\sigma_{i} e^{-j2k\cdot R_{i}}\Bigg|^{2},
   \label{RCS_eq2}
\end{equation}
where $\sigma_i$ is the RCS of the $i$th scattering point on the complex target, which is a distance $R_i$ from the radar. Given that $R_i$ is a random variable, the total RCS ($\sigma$) of the complex target will be random, with some probability density $p(\sigma)$. 


For many complex targets such as small drones, the measured RCS data can be used to determine the best probability density model that describes the RCS fluctuations of the drone. In this study, the measured RCS data from three different small drones, obtained from the compact range anechoic chamber measurement, will be fitted to several statistical distributions. The family of statistical distribution investigated includes log-normal, Rayleigh, and generalized extreme value.
 
 In this study, the Akaike information criterion (AIC) is used for RCS model selection. The AIC criterion tries to measure the fit of a model to data while penalizing the model complexity. For a given candidate model~\rm{M}, the AIC is computed as follows: 
\begin{equation}
   \rm{AIC}(M)=-2\sum_{i=1}^{N}log \mathcal{L}(\boldsymbol {\hat\theta}|\sigma_i) + 2K,
   \label{RCS_eq4}
\end{equation}
where $N$ is the number of samples (or data points) in the measured RCS data, $ \boldsymbol{\hat{\theta}}$ is the maximum likelihood estimate of the parameter vector of the model, and $K$ is the number of estimated parameters in the model. In (\ref{RCS_eq4}), $\mathcal{L}(\boldsymbol {\hat\theta}|\sigma_i)$ measures the log-likelihood of the model at its maximum point. The best RCS model for a given drone is the one with the smallest AIC.


 
\subsection{Measurements in Compact-Range Anechoic Chamber}
The compact range anechoic chamber enables the creation of plane wave illumination in an indoor environment at microwave frequencies. Plane-wave target illumination fulfills the far-field requirement for accurate RCS measurement.  Fig.~\ref{EXPERIMENTAL_setup} shows the schematic of the compact range anechoic chamber used in this experimental study. The main functional components of the chamber include a 20 foot collimating parabolic reflector, Keysight E8362B programmable vector network analyzer (VNA), two H-1498 broadband horn antennas (transmit (TX) and receive (RX) antennas), high performance pyramidal and wedge absorbers, a Styrofoam turntable and a windows computer system for automation and control.

During the measurement, the target drone is placed on the Styrofoam turntable which is 6 feet away from the antennas. The turntable, controlled by a stepper motor, rotates the target through the azimuth plane $\phi\in[0\degree, 360\degree]$ with a 2$\degree$ increment. For each look angle, continuous wave signals, centered at the test frequency, are generated in the VNA are transmitted through the transmit horn antenna (TX) which is located at the focus of the parabolic reflector. The curvature and smoothness of the parabolic reflector ensure that the reflected waves are collimated to simulate far-field conditions at a relatively short distance. The incident plane waves are scattered by the target. The receive horn antenna (RX), which is connected to the input port of the VNA, captures the scattered signals. The scattered signal power is processed to measure the RCS of the small drone. Fig.~\ref{ANECHOIC_CHAMBER} shows a typical measurement scenario for the Trimble zx5 drone.

\subsection{Post Processing}

Post-processing is done in MATLAB. Three major post-processing operations are performed on the received signal. These operations are software range gating, background subtraction, and calibration. Software gating is required to isolate the backscattered signals from other unwanted spurious signals such as leakage in the transceiver system and coupling between the horn antennas. The software range gating involves transforming the captured signals from frequency domain to time domain using inverse Fourier transform. Afterward, a Tukey window (tapered cosine window) function which corresponds to the target gate is used to filter out unwanted time-domain responses from the captured signal. Also, we capture the background (without a drone on the turntable) to remove the reflections of the empty chamber from the captured data.

For measurement calibrations, we use a perfectly electrical conducting (PEC) 12~inch metallic sphere shown in Fig.~\ref{caliberation_sphere}. The 12 inch PEC sphere is used for experimental calibrations because the far-field monostatic RCS of a PEC sphere can be predicted using the closed-form analytical expression given by~\cite{balanis1999advanced}:
\begin{equation}
\begin{aligned}
\label{RCS_sphere_equ}
     \sigma_{\rm{sphere~exact}} &=\frac{\lambda^2}{4\pi}\Bigg|\sum_{n=1}^{\infty}\frac{(-1)^n(2n+1)}{\hat{H}_n^{(2)'}(ka){\hat{H}_n^{(2)}(ka)}}\Bigg|^2,\\
\end{aligned}
\end{equation}
where $k=\frac{2\pi}{\lambda}$ is the wavenumber, $a$ is the radius of the sphere, ${\hat{H}_n^{(2)}(ka)}$ and ${\hat{H}_n^{(2)'}}$ are the spherical Hankel function of the second kind of order $n$ and its derivative, respectively. From (\ref{RCS_sphere_equ}), we can show that the backscattered RCS of a PEC sphere is a function of its circumference measured in wavelength ($\frac{2\pi a}{\lambda}$). In the Rayleigh region, the circumference (or size) of the PEC sphere is small relative to the wavelength of the transmitted signal ($\frac{2\pi a}{\lambda}\ll1$). As a result, in the Rayleigh region, the theoretical RCS of the ideal PEC sphere can be approximated as $\frac{9\lambda^2}{4\pi}(ka)^6$. On the other hand, in the optical region, the circumference of the PEC sphere is far larger than the wavelength of the transmitted signal ($a>2\lambda$). As a result, in the optical region, the theoretical RCS of the ideal PEC sphere is independent of $\lambda$ and can be approximated by a constant given as $\pi a^2$. 

\begin{figure}[t!]
 \center
 \includegraphics[scale=0.45]{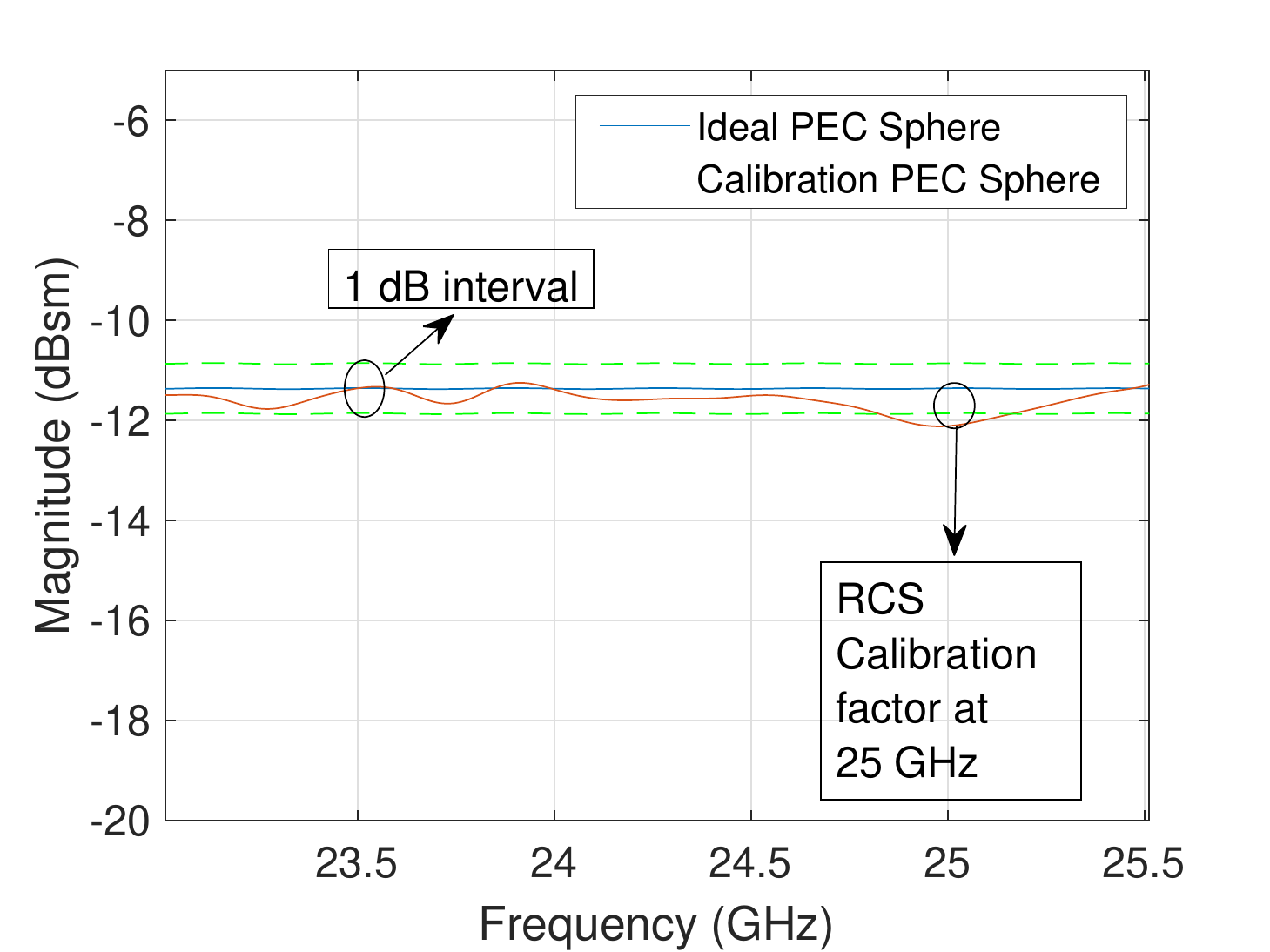}
\caption{The simulated RCS of an ideal PEC sphere versus measured RCS of the calibration PEC sphere at frequencies around 25 GHz (optical far-field region). The difference between the RCS values of the ideal sphere and the calibration sphere is used to calibrate the measured RCS of the drones at 25 GHz.}
\label{Fig:calibration}
\end{figure}

\begin{figure}{}
\center{
\begin{subfigure}[]{\includegraphics[width=0.27\linewidth]{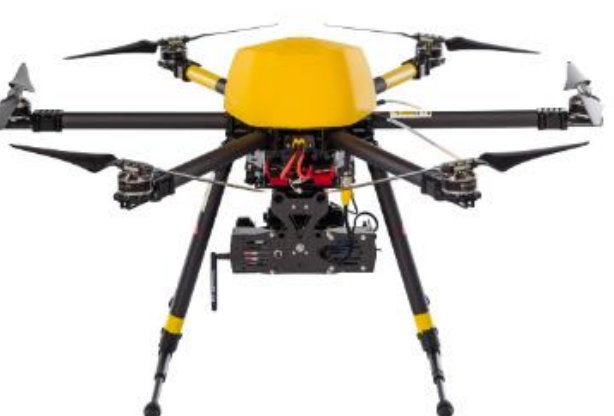}\label{trimble_UAV}}
\end{subfigure}
\begin{subfigure}[]{\includegraphics[width=0.27\linewidth]{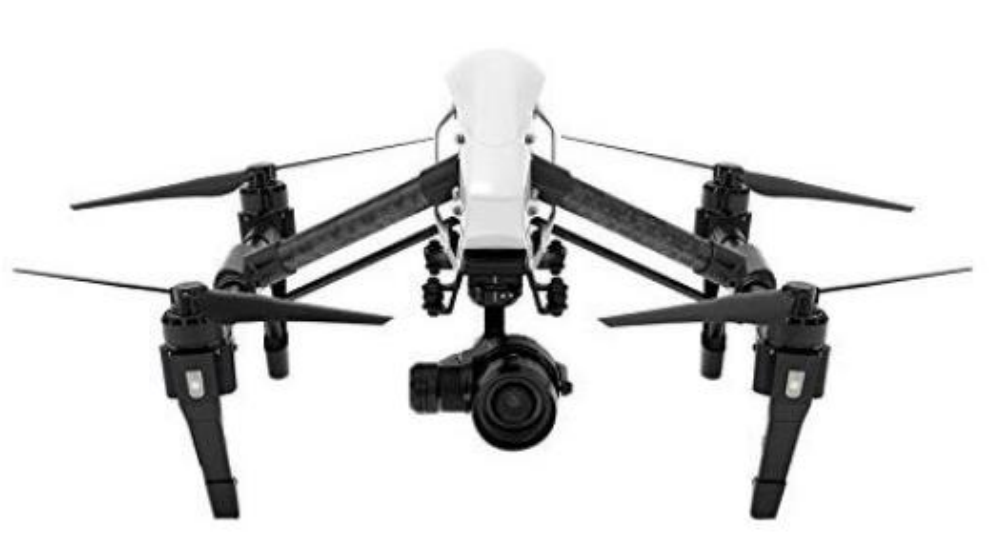}\label{DJI_inspire}}
\end{subfigure}
\begin{subfigure}[]{\includegraphics[width=0.27\linewidth]{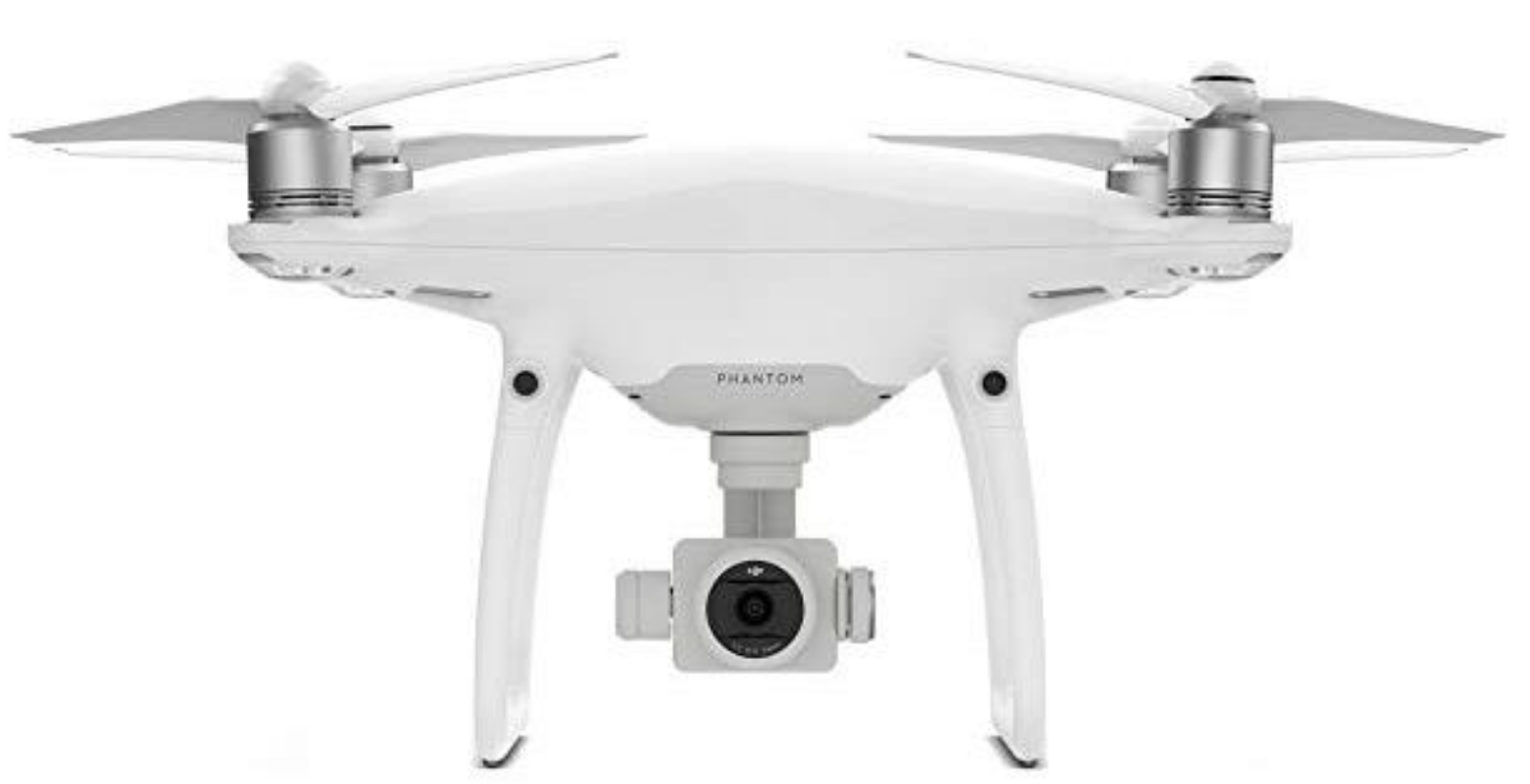}\label{DJI_phantom4pro}}
\end{subfigure}
 \caption{{Small drones considered: (a) Trimble zx5, (b) DJI Inspire 1 Pro, (c) DJI Phantom 4 Pro (DJI P4 Pro).}}
 \label{SMALL_DRONES}}
 \end{figure}

Moreso, between the Rayleigh and the optical regions is the Mie region. The Mie region is also known as the resonance region because it is characterized by continuous perturbation in the RCS of the sphere. However, since the RCS of an ideal PEC sphere is approximately constant in the optical region, the 12 inch PEC spheres with a known theoretical RCS of about -11.37 dBsm (in the optical region) is used to calibrate the drone RCS measurement in the compact range anechoic chamber. Fig.~\ref{Fig:calibration} shows the simulated RCS of an ideal PEC sphere and the measured RCS of the actual calibration PEC sphere at 25 GHz. The difference in RCS at 25 GHz is used for calibrating the drone RCS measurement at that frequency. A similar calibration operation is performed for the drone RCS measurement at 15 GHz. Fig.~\ref{SMALL_DRONES} shows the three small drones used in this experimental study. These are popular commercial drones with body frames and propellers made of low reflective materials like carbon fiber and plastic.

\begin{figure*}{}
\center{
\begin{subfigure}[]{\includegraphics[width=0.29\linewidth]{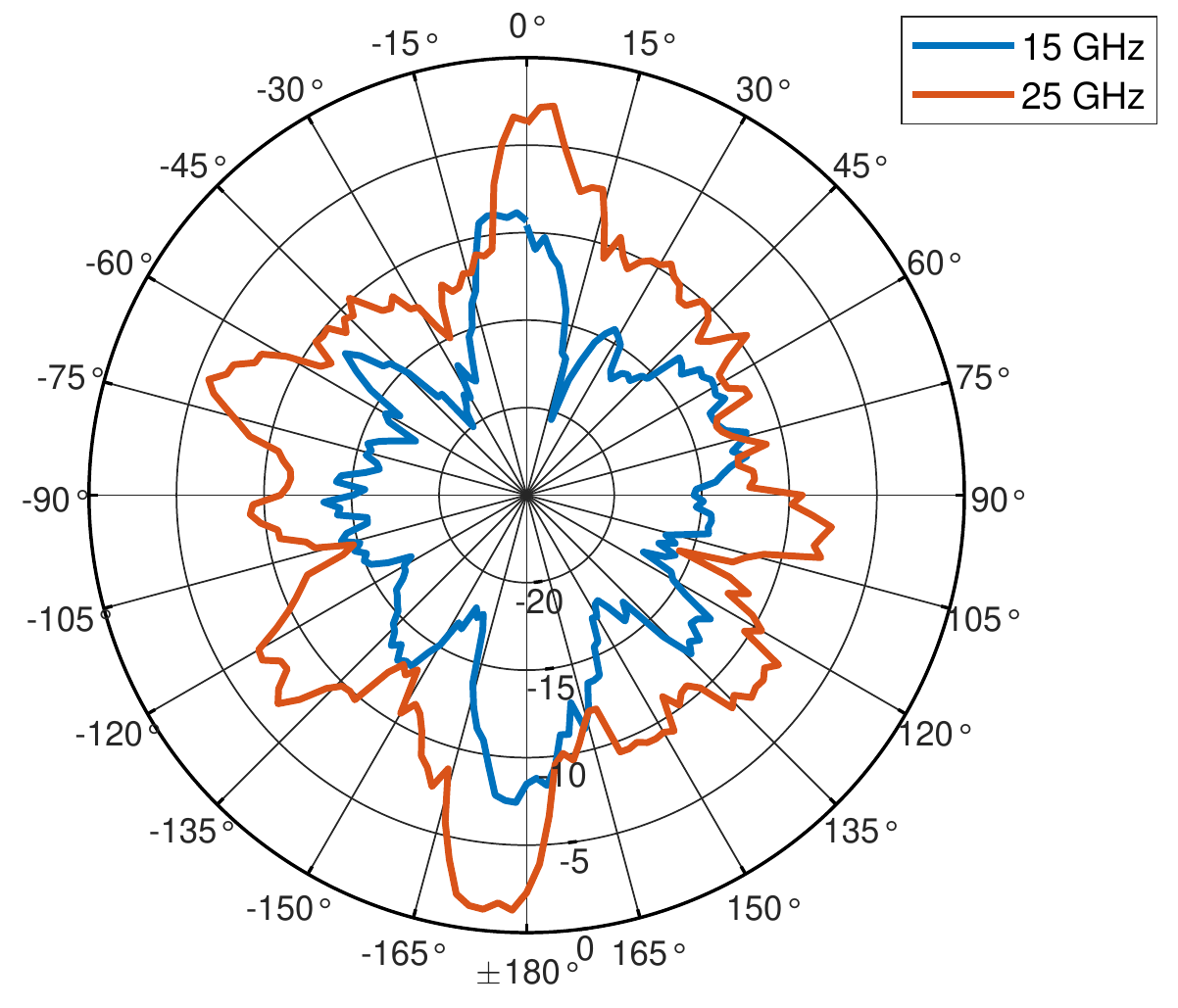}\label{trimbler_RCS_UAV}}
\end{subfigure}
\begin{subfigure}[]{\includegraphics[width=0.29\linewidth]{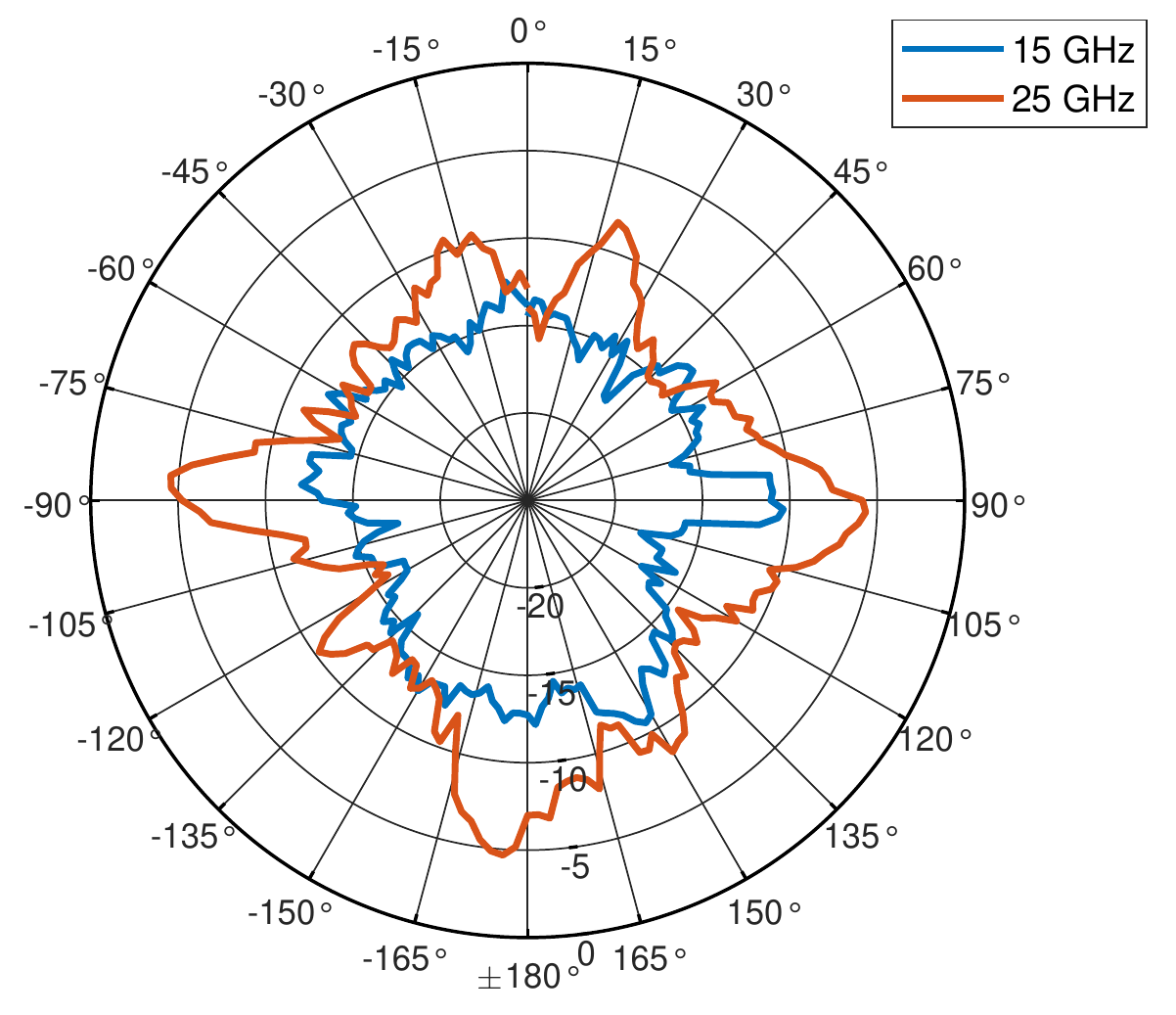}\label{inspire_RCS_UAV}}
\end{subfigure}
\begin{subfigure}[]{\includegraphics[width=0.29\linewidth]{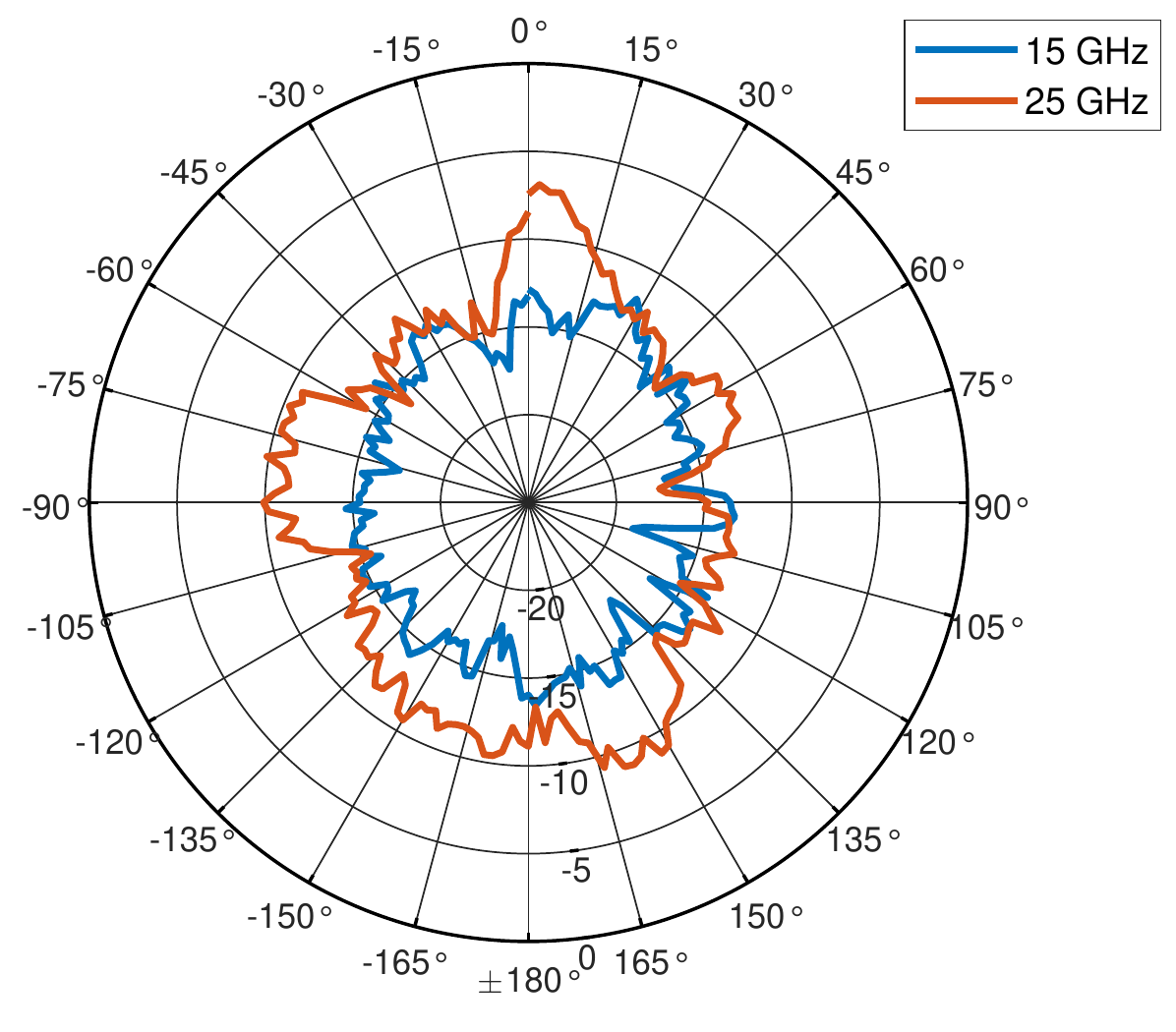}\label{phantom_RCS_UAV}}
\end{subfigure}
 \caption{{The measured RCS (dBsm) versus azimuth angles ($\phi\in[0\degree, 360\degree]$) for the small drones: (a) Trimble zx5, (b) DJI Inspire 1, (c) DJI Phantom 4 Pro.}}
 \label{RCS_POLAR_PLOT}}
 \end{figure*}
\begin{table}[t]
\centering
\caption{The AIC test score for the three probability models that are used to fit the compact-range RCS data of the small drones. Model 1: Log-normal, Model 2: Rayleigh, and Model 3: Generalized extreme value. The average RCS value ($\bar{\sigma}$) is also provided. }
\label{AIC_table}
\begin{tabular}{|p{0.58cm}|c|c|c|c|c|}
\hline
Freq.& Drone & $\bar{\sigma}$ &\multicolumn{3}{c}{AIC Test Score}\\ 
 (GHz)& & (dBsm)  & Model 1 & Model 2 & Model 3\\
\hline
\multirow{3}{*}{\text{15}}& Trimble zx5 & -14.39 & 327.30 & 409.52 & 318.12\\
& DJI Inspire 1 & -14.24 & 158.70& 218.97 &160.60\\
& DJI P4 Pro & -15.03 & 300.11 & 414.10 & 301.70\\
\hline
\multirow{3}{*}{\text{25}}&Trimble zx5 &-9.64 & 754.23 &  931.10 & 726.30\\
& DJI Inspire 1 & -11.09 & 610.04 & 689.27 & 596.14\\
& DJI P4 Pro & -12.40 & 388.98 &  434.52 &387.56\\
\hline
\end{tabular}
\end{table}
\section{Results and Discussion}\label{result_DISCUSSION}
 Fig.~\ref{RCS_POLAR_PLOT} shows in polar plot the measured RCS (dBsm) versus azimuth angle for each of the small drones. For the three small drones considered in this study, the RCS value in all azimuth plane is less than 0~dBsm. At 15 GHz, the average RCS of the Trimble zx5, DJI Inspire 1, and DJI Phantom 4 Pro (DJI P4 Pro) are -14.39 dBsm (0.0364 $m^2$), -14.24 dBsm (0.0377 $m^2$), and -15.03 dBsm (0.0314 $m^2$), respectively. At 25 GHz, the average RCS of the Trimble zx5, DJI Inspire 1, and DJI P4 Pro are -9.64 dBsm (0.1087 $m^2$), -11.09 dBsm (0.0778 $m^2$), and -12.40 (0.0576 $m^2$), respectively. These low RCS values are primarily due to the low reflective materials (plastic and carbon fiber) used in the design of these small drones. The measurement results show that the RCS of the Trimble zx5 drone, a relatively bigger drone, is larger than the RCS of the other two smaller drones. 
 
 Moreover, for all three drones, the average RCS is higher at 25 GHz than at 15 GHz. This observation agrees with the conclusion in~\cite{nakamura2018ultra} which compares the RCS of a small drone measured at 24, 26, 60, and 79 GHz. The author observed that for the small drone, the RCS at the higher frequency was larger than the RCS measured at the lower frequency. Besides, in~\cite{nakamura2018ultra}, the average RCS of DJI Phantom 3 (DJI P3 Pro) measured at 24 GHz is given as -13~dBsm. The DJI P3~Pro has a similar shape, material property, and size as the DJI P4~Pro which is one of the small drones investigated in this study. Therefore, the average RCS value of the DJI P3 Pro measured at 24 GHz in~\cite{nakamura2018ultra} is close to the average RCS of the DJI P4 Pro which we have measured at 25 GHz as -12.40 dBsm. 
 
Since the RCS measured at 25 GHz radar is higher, this radar radar would be able to identify the radar signature of a small drone better than a 15 GHz radar. This is because the very small wavelength ($\lambda=1.2$~cm) of the 25 GHz radar enhances the effects of diffraction and scattering by sharp edges and corners on the drones.
Consequently, a 25 GHz radar would be more suitable in designing an automatic target recognition (ATR) system for identifying small drones. Also, Fig.~\ref{RCS_POLAR_PLOT} shows that for each drone, the RCS values are higher in specific directions. Therefore, recognizing these dominant scattering centers can help us distinguish small drones from other airborne objects.

Table~\ref{AIC_table} shows the result of the model selection based on the AIC test score. For each test frequency, the best model is the one with the lowest AIC score. From the Table~\ref{AIC_table}, we see that in most cases, the extreme value distribution fits best with the RCS data. The exceptions occur for the RCS data of DJI Inspire 1 and DJI Phantom 4 Pro measured at 15 GHz where the log-normal distribution is most suitable. Therefore, using the appropriate scattering model, we can design an effective model-based statistical technique for radar-based detection of small drones. 

\section{Conclusion}\label{conclusion_sec}
The paper describes a compact-range approach for measuring the RCS of different small drones at 15 GHz and 25 GHz. We present some preliminary results which show that the RCS or radar signature of small drones is unique. This knowledge can be used to identify different small drones. Moreover, we showed how the RCS data can be modeled using different parametric models. Future work will investigate how these parametric models can be used to develop model-based statistical techniques for detection and identification of small drones.

\section{Acknowledgment}

This work has been supported by the NASA grant NNX17AJ94A. Also, the authors are grateful to Mr. Kenneth Ayotte and the management of the Ohio State University Electroscience Laboratory for helping out with the experiments.

\bibliography{IEEEabrv,references}

\begin{thebibliography}{10}
\providecommand{\url}[1]{#1}
\csname url@rmstyle\endcsname
\providecommand{\newblock}{\relax}
\providecommand{\bibinfo}[2]{#2}
\providecommand\BIBentrySTDinterwordspacing{\spaceskip=0pt\relax}
\providecommand\BIBentryALTinterwordstretchfactor{4}
\providecommand\BIBentryALTinterwordspacing{\spaceskip=\fontdimen2\font plus
\BIBentryALTinterwordstretchfactor\fontdimen3\font minus
  \fontdimen4\font\relax}
\providecommand\BIBforeignlanguage[2]{{%
\expandafter\ifx\csname l@#1\endcsname\relax
\typeout{** WARNING: IEEEtran.bst: No hyphenation pattern has been}%
\typeout{** loaded for the language `#1'. Using the pattern for}%
\typeout{** the default language instead.}%
\else
\language=\csname l@#1\endcsname
\fi
#2}}

\bibitem{shakhatreh2018unmanned}
H.~Shakhatreh, A.~H. Sawalmeh, A.~Al-Fuqaha, Z.~Dou, E.~Almaita, I.~Khalil,
  N.~S. Othman, A.~Khreishah, and M.~Guizani, ``Unmanned aerial vehicles
  ({UAVs}): A survey on civil applications and key research challenges,''
  \emph{IEEE Access}, vol.~7, pp. 48\,572--48\,634, Apr. 2019.

\bibitem{ezuma2019micro}
M.~Ezuma, F.~Erden, C.~K. Anjinappa, O.~Ozdemir, and I.~Guvenc, ``Micro-{UAV}
  detection and classification from {RF} fingerprints using machine learning
  techniques,'' in \emph{Proc. IEEE Aerosp. Conf., Big Sky, Montana}, Mar.
  2019.

\bibitem{guvenc2018detection}
I.~Guvenc, F.~Koohifar, S.~Singh, M.~L. Sichitiu, and D.~Matolak, ``Detection,
  tracking, and interdiction for amateur drones,'' \emph{{IEEE} Commun. Mag.},
  vol.~56, no.~4, pp. 75--81, Apr. 2018.

\bibitem{MartinsDrone}
M.~Ezuma, O.~Ozdemir, C.~Kumar, W.~A. Gulzar, and I.~Guvenc, ``{Micro-UAV}
  detection with a low-grazing angle millimeter wave radar,'' in \emph{Proc.
  IEEE Radio Wireless Week (RWW) Conf.}, Orlando, FL, Jan. 2019.

\bibitem{White_House}
\BIBentryALTinterwordspacing
M.~Schmidt and M.~Shear, ``A drone, too small for radar to detect, rattles the
  {White House},'' 2015. [Online]. Available:
  \url{https://www.nytimes.com/2015/01/27/us/white-house-drone.html}
\BIBentrySTDinterwordspacing

\bibitem{rahman2018flight}
S.~Rahman and D.~A. Robertson, ``In-flight {RCS} measurements of drones and
  birds at k-band and w-band,'' \emph{IET Radar, Sonar \& Navigation}, vol.~13,
  no.~2, pp. 300--309, Sept. 2018.

\bibitem{gong2019interference}
J.~Gong, J.~Yan, D.~Li, D.~Kong, and H.~Hu, ``Interference of radar detection
  of drones by birds,'' \emph{Progress In Electromagnetics Research}, vol.~81,
  pp. 1--11, Apr. 2019.

\bibitem{nakamura2017characteristics}
R.~Nakamura and H.~Hadama, ``Characteristics of ultra-wideband radar echoes
  from a drone,'' \emph{IEICE Communications Express}, vol.~6, no.~9, pp.
  530--534, June 2017.

\bibitem{roding2017fully}
M.~R{\"o}ding, G.~Sommerkorn, S.~H{\"a}fner, R.~M{\"u}ller, R.~S. Thom{\"a},
  J.~Goerlich, and K.~Garhammer, ``Fully polarimetric wideband rcs measurements
  for small drones,'' in \emph{Proc. European Conf. Antennas Propag. (EUCAP)},
  Paris, France, Mar. 2017, pp. 3926--3930.

\bibitem{ancortek_radar}
\BIBentryALTinterwordspacing
 [Online]. Available: \url{https://https://ancortek.com/productoverview}
\BIBentrySTDinterwordspacing

\bibitem{fortem_radar}
\BIBentryALTinterwordspacing
 [Online]. Available: \url{https://fortemtech.com/}
\BIBentrySTDinterwordspacing

\bibitem{knott2004radar}
E.~F. Knott, J.~F. Schaeffer, and M.~T. Tulley, \emph{Radar cross
  section}.\hskip 1em plus 0.5em minus 0.4em\relax Raleigh, NC: SciTech
  Publishing, 2004.

\bibitem{balanis1999advanced}
C.~A. Balanis, \emph{Advanced {E}ngineering {E}lectromagnetics}, 2nd~ed.\hskip
  1em plus 0.5em minus 0.4em\relax Hoboken, NJ: John Wiley \& Sons, 2012.

\bibitem{nakamura2018ultra}
R.~Nakamura, H.~Hadama, and A.~Kajiwara, ``Ultra-wideband radar reflectivity of
  a drone in millimeter wave band,'' \emph{IEICE Communications Express},
  vol.~7, no.~9, pp. 341--346, 2018.

\end{thebibliography}
\bibliographystyle{IEEEtran}
\end{document}